# Electron vacancy-level dependent hybrid photoionization of F$^-$@C$_{60}^+$ molecule

Esam Ali,[1, *] Taylor O'Brien,[1] Andrew Dennis,[1] Mohamed El-Amine
Madjet,[1, 2] Steven T. Manson,[3] and Himadri S. Chakraborty[1, †]

[1]*Department of Natural Sciences, D.L. Hubbard Center for Innovation,
Northwest Missouri State University, Maryville, Missouri 64468, USA*
[2]*Max-Planck-Institut für Physik komplexer Systeme, Nöthnitzer Straße 38, 01187 Dresden, Germany*
[3]*Department of Physics and Astronomy, Georgia State University, Atlanta, Georgia 30303, USA*

(Dated: June 4, 2021)

Our previous studies [J. Phys. B **53**, 125101 (2020); Euro. Phys. J. D **74**, 191 (2020)] have predicted that the atom-fullerene hybrid photoionization properties for X = Cl, Br and I endohedrally confined in C$_{60}$ are different before and after an electron transfers from C$_{60}$ to the halogen. It was further found as a rule that the ionization dynamics is insensitive to the C$_{60}$ level the electron originates from to produce X$^-$@C$_{60}^+$. In the current study, we report an exception to this rule in F@C$_{60}$. It is found that when the electron vacancy is situated in the C$_{60}$ level that participates in the hybridization in F$^-$@C$_{60}^+$, the mixing becomes dramatically large leading to strong modifications in the photoionization of the hybrid levels. But when the vacancy is at any other pure level of C$_{60}$, the level-invariance is retained showing weak hybridization. Even though this case of F@C$_{60}$ is an anomaly in the halogen@C$_{60}$ series, the phenomenon can be more general and can occur with compounds of other atoms caged in a variety of fullerenes. In addition, possible experimental studies are suggested to benchmark the present results.

## I. INTRODUCTION

Progresses in synthesis and isolation techniques [1] of endohedral fullerenes, an atom taken hostage in a fullerene cavity, or endofullerenes for short, present opportunities for their experimental and motivations for their theoretical studies. One attraction for such studies arises from the interesting symmetry and great stability of these systems of natural molecular traps. The other is the fact that a vast range of existing and potential technologies, encompassing photovoltaics [2], superconductivity [3], quantum computations [4, 5], molecular device [6] and bio-medics [7], uses materials that have endofullerenes at their core. Such technologies and their further developments may find underpinnings of success from the information obtained *via* fundamental spectroscopic studies of these molecules.

In order to study the ionizing response of vapor-phase endofullerenes to UV electromagnetic radiations, a series of measurements [8–10] was conducted using merged beam techniques at the Berkeley Advanced Light Source. With the increase of sample production rates, there are future possibilities of performing photoelectron spectroscopic experiments [11] on these systems. A swath of theoretical studies at various levels of approximation on the photoionization of endofullerenes with closed shell atoms also exists [12, 13]. In particular, hybridization of highlying orbitals of the central atom with C$_{60}$ and explorations of the photoionization properties of these hybrid levels have regularly been predicted [14–18]. Hybrid levels are unique, for they expose interference features from the atom-fullerene coherent ionization which are absent in the isolated systems.

Endofullerenes of open-shell atoms, on the other hand, may also have interesting applied relevance [19]. At a rather exotic level, N@C$_{60}$ exhibited uniquely long spin relaxation times driven by the confinement [20], P@C$_{60}$ displayed an enhancement in hyperfine coupling of the phosphorous' unpaired electrons with its nucleus [21], and muonium@C$_{60}$ indicated a diminution of the hyperfine interaction between the positively-charged muon and the unpaired electrons [22].

Computations of nonlinear optical response were done on endofullerene dimers where C$_{60}$ is used as one monomer [23]. Very recently, we performed theoretical photoionization studies of isolated halogen endofullerenes Cl@C$_{60}$ [24], Br@C$_{60}$ and I@C$_{60}$ [25]. A halogen atom with an outer-shell electron vacancy (hole) can be reactive and can readily capture an electron from C$_{60}$ to acquire a stable configuration. However, this transfer will need energy, since the electron affinity level energies of halogens are higher than the binding energy of C$_{60}$ HOMO level. Of course, there will be extra binding due to the electrostatic attraction between F$^-$ and C$_{60}^+$. However, some such stable configurations with the electron coming from deeper C$_{60}$ levels may actually be metastable due to the need for larger transition energies. In any case, the realistic ground state of the compound may be a mixture of configurations, before and after the electron transfer. Therefore, knowledge of photoionization properties of all molecular states can be useful to reveal the actual configuration of the molecule by photoelectron spectroscopic experiments. Furthermore, multiple metastable states are known to form in other processes, such as, in slow electron collisions leading to the formation of negative ions of fullerene molecules [26]. Therefore, a knowledge repository of endofullerenes with various stable and metastable electronic structures accessed by photoionizing the atom-fullerene hybrid levels can have cross-topical relevance as well.

* eali@nwmissouri.edu
† himadri@nwmissouri.edu

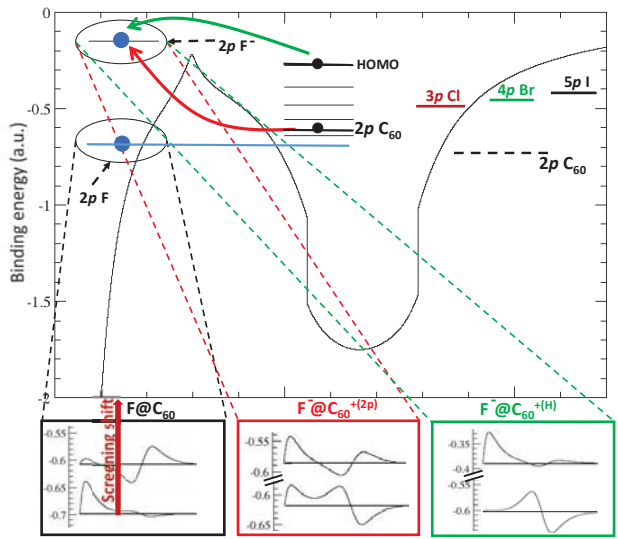

FIG. 1. (Color online) A schematic that illustrates the $C_{60}$ vacancy-level dependence in the ground state hybridization of the compound (see text).

It was found in our previous studies [24, 25] that the photoionization properties of the configurations of $Cl^-@C_{60}^+$, $Br^-@C_{60}^+$ and $I^-@C_{60}^+$ were not sensitive to the particular $C_{60}$ level where the hole is located. This is because, as shown in Fig. 1, the valence $np$ levels of the isolated X = Cl, Br, and I are energetically above the $2p$ level of empty $C_{60}$ (in the jellium model) that participates in the hybridization. When an electron in $X@C_{60}$ is captured by the central atom, the resulting screening effect shifts the atomic $np$ level even higher. So the separation of interacting levels gets so large that it becomes irrelevant whether the electron that moved was a pure $C_{60}$ electron or a hybrid electron of $X@C_{60}$. Thus, the modification of the hybridization after the transfer becomes effectively insensitive to the vacancy level of $X^-@C_{60}^+$. The situation, however, is very different if the free atomic level locates, on the other hand, below $2pC_{60}$ as is the case for the smallest halogen F (see Fig. 1). In forming $F^-@C_{60}^+$, the screening from the electron transfer will cause $2pF$ to rise and move toward $2pC_{60}$ from below it (Fig. 1, bottom left panel). This will create the condition of a reduced level separation, instead of the increase as happens for other halogens. The effect may sensitize the hybridization mechanism. For example, the transfer of a hybrid versus a pure $C_{60}$ electron could make a difference. These transfers are illustrated in Fig. 1, which also displays the resulting hybrid energy-levels and radial wavefunctions, respectively, in the middle and the right bottom panels. As we demonstrate in this paper, the hybridization indeed becomes dramatically higher when the vacancy is generated in the hybrid-active level of $F@C_{60}$

(bottom middle panel). To illustrate the effect of the vacancy at a pure (hybrid-passive) level which greatly minimizes the hybridization, HOMO $C_{60}$ is considered (bottom right panel), since the result is found practically identical for any pure $C_{60}$ vacancy. Strong implications of this phenomenon on the hybrid level photoionization of $F^-@C_{60}^+$ will be presented by comparing cross sections calculated with the hole in different $C_{60}^+$ levels.

## II. THEORETICAL MODEL IN BRIEF

The details of the theoretical schemes are described in Ref. [15] and more recently in Ref. [24]. Choosing the photon polarization along the $z$-axis, the photoionization dipole transition cross section in a linear response approximation of time-dependent density-functional theory (DFT) is given by,

$$\sigma_{n\ell \to k\ell'} \sim |\langle \psi_{\mathbf{k}\ell'}|z + \delta V|\phi_{n\ell}\rangle|^2. \quad (1)$$

Here $\mathbf{k}$ is the momentum of the continuum electron, $z$ is the one-body dipole operator, $\phi_{nl}$ is the single electron bound wavefunction of the target level, and $\psi_{\mathbf{k}l'}$ is the respective outgoing dipole-allowed continuum wavefunction, with $l' = l \pm 1$. $\delta V$ represents the complex induced potential that accounts for electron correlations within the linear response of the electrons to the photon field. The computation of $\delta V$ involves determining photon energy dependent induced change in the electron density to be obtained by varying the ground state potential with respect to the ground state electron density as described in Ref. [27].

We model the bound and continuum states, and the ground state potential, using the independent particle DFT approximation that utilizes the Leeuwen-Baerends (LB) exchange-correlation functional [28]. This functional involves the gradient of the electron density in the scheme described earlier [29]. We chose the spherical frame with F or $F^-$ situating at the center of $C_{60}$. The polarization interaction of $F^-$ may induce some offset in its position from the center. However, a DFT simulation with Born-Oppenheimer molecular dynamics found $F^-$ oscillation within neutral $C_{60}$ to be rather small [30] and a relatively weak effect on the photoionization process from such small offset was predicted [31]. Earlier studies also showed small effects of the cage polarization except very close to the ionization threshold [32].

A core of 60 $C^{4+}$ ions for $C_{60}$ is constructed by smearing the total positive charge over a spherical jellium shell with known molecular radius $R = 6.70$ a.u. $(3.54Å)$ [11] and thickness $\Delta$. The Kohn-Sham equations for the system of 240 $C_{60}$ electrons (four valence $2s^2 2p^2$ electrons from each carbon atom), $plus$ all electrons of the central atom/ion, are then solved self-consistently. The values of $\Delta$ and a pseudo potential used are determined both by requiring charge neutrality and obtaining the experimental value [33] of the first ionization threshold of $C_{60}$. $\Delta = 2.46$ a.u. $(1.30Å)$ thus obtained closely agree with the value extracted from measurements [11, 34]. Within this framework, we also selectively omit either F (F⁻) or $C_{60}$ ($C_{60}^+$) to obtain the corresponding empty $C_{60}$ ($C_{60}^+$) and free F (F⁻) results. Parametric optimization of LB functional followed the scheme utilized in the previous work [25] to reproduce the values of the ionization potential and the electron affinity of F from the NIST database within 20%.

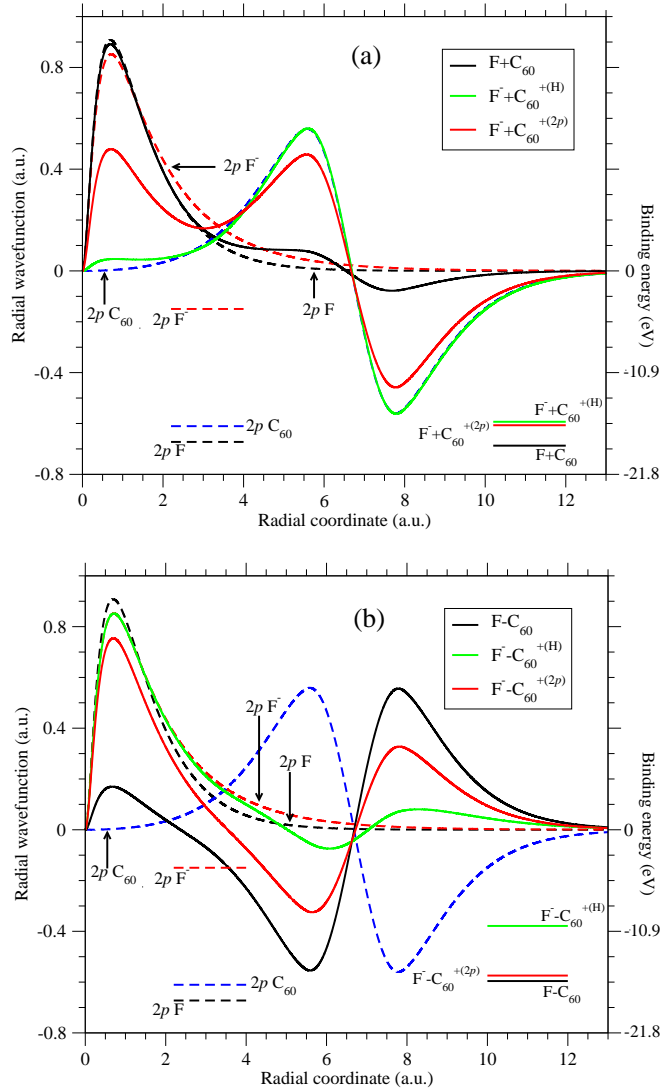

FIG. 2. (Color online) Radial wavefunctions (left vertical scale) and level energies (right vertical scale) of the symmetric (a) and antisymmetric (b) hybrid states of F@$C_{60}$, F⁻@$C_{60}^{+(H)}$, and F⁻@$C_{60}^{+(2p)}$ are displayed. Wavefunctions and energies of $2p$F (F⁻) and $2p$$C_{60}$ are also shown for reference.

## III. RESULTS AND DISCUSSION

### A. Ground state hybridization versus vacancy levels

In an endofullerene system, an eigenstate of the free atom can only couple with an eigenstate of the empty $C_{60}$ of the same angular momentum symmetry (due to the orthogonality of spherical wavefunctions) to produce a pair of hybrid eigenstates of the composite system. For F@$C_{60}$ this occurs between valence $2p$F and $2p$$C_{60}$ state to produce symmetric and antisymmetric hybrid states of F@$C_{60}$ as,

$$|F \pm C_{60}\rangle = \eta_{\pm}|\phi_{2pF}\rangle \pm \eta_{\mp}|\phi_{2pC_{60}}\rangle \quad (2)$$

for F@$C_{60}$, and the same for F⁻@$C_{60}^+$ with F and $C_{60}$ replaced by F⁻ and $C_{60}^+$. Here $\eta_+ = \sqrt{\alpha}$ and $\eta_- = \sqrt{1-\alpha}$, where $\alpha$ is the mixing parameter that renders the hybrid states orthonormal. We use the hydrogenic Coulomb notation of labeling for F and F⁻, such that the corresponding $2p$ wavefunctions are radially nodeless. The standard harmonic oscillator notion is used for $C_{60}$ and $C_{60}^+$ where their $2p$ wavefunctions have one radial node. The radial component of these hybrid wavefunctions are, respectively, shown in Figs. 2(a) and (b). In general, the hybridization is favored when the energy separation between the reacting levels decreases and their wavefunction overlap increases; these energies and wavefunctions are also shown in Fig. 2. While the free $2p$F level is just below empty $2p$$C_{60}$, their wavefunction overlap is small owing to the small size of F leading to a weak hybridization in F@$C_{60}$. As seen, the deeper symmetric hybrid F+$C_{60}$ [Fig. 2(a)] still retains significant F character, while the weaker-bound antisymmetric hybrid F-$C_{60}$ [Fig. 2(b)] continues to remain largely $C_{60}$-like. On the other hand, the hybrid states of F⁻@$C_{60}^{+(H)}$ with the vacancy at the HOMO $C_{60}$ level (which does not participate in hybridization), also plotted in Fig. 2, present the opposite picture: the significantly weaker bound antisymmetric hybrid [Fig. 2(b)] now becomes F⁻-type, while the symmetric hybrid [Fig. 2(a)] is of $C_{60}$-type. We find practically the same degree of mixing when the vacancy is at any other hybrid-passive $C_{60}$ level, like HOMO, that remains pure in the compound. Note the following in going from F@$C_{60}$ to F⁻@$C_{60}^{+(H)}$ in Fig. 2: (i) The binding energy of F⁻-dominant hybrid level is significantly lower than that of the F-dominant hybrid level which is because of the stronger screening of the F nucleus due to the extra electron that was transferred to the atomic zone. However, (ii) the cage-dominant hybrids remain energetically close primarily due to the fact that the relocation of a single electron hardly affects the screening of the cage states owing to the large electron population of $C_{60}$. In effect, therefore, the F-dominated level swings past the almost stagnant cage-dominated level as the molecular configuration switches from F@$C_{60}$ to F⁻@$C_{60}^{+(H)}$. However, if we choose the configuration F⁻@$C_{60}^{+(2p)}$, that is the va-



cancy in the hybrid-active $2pC_{60}$ level, the hybridization becomes almost 50-50 complete with the hybrid level energies being very close, as Fig. 2 further displays. The bottom three panels of Fig. 1 together summarize the essence of the total phenomenon.

To understand the underlying physics that leads to the enhancement of the hybridization in $F^-@C_{60}^+$ when the vacancy is in $2pC_{60}$ versus in HOMO $C_{60}$, we use a qualitative picture. Although the calculation is done by placing a $F^-$ ion in a $C_{60}^+$ cage directly, it is convenient to consider a two step model. First consider the ground state of $F@C_{60}$. Then we use the levels of $F@C_{60}$, including its hybrid levels, as the basis to analyze the hybridization after an electron moves from $C_{60}$ to $F$. This electron transfer is considered selectively from two levels: (i) from HOMO of $F@C_{60}$ which is also HOMO of $C_{60}$ and a pure $C_{60}$ level and (ii) from $F$-$C_{60}$ which is a hybrid level of $F@C_{60}$ but of dominantly cage-type and energetically shallower than the other hybrid $F+C_{60}$ which is dominantly $F$-type (Fig. 2). In the first case it is a "whole" electron that transfers to the F-zone and causes a "complete" screening of the atomic nucleus. This causes the F-dominant hybrid of $F@C_{60}$ to energetically shift upward, cross the practically unmoved cage-dominant hybrid and move far above to result in yet another weak hybridization in $F^-@C_{60}^{+(H)}$, albeit a symmetry reversal as noted above and seen in Fig. 2. This effect will be quite general as long as the transferring electron originates from any of the pure $C_{60}$ states of $F@C_{60}$, even though the total energy of resulting configurations will be different in each case. However, when the electron comes from the hybrid $F$-$C_{60}$ level, effectively a "partial" electron transfers, inducing a rather "incomplete" screening effect. Consequently, the F-dominant level will still up-shift, but under a weaker screening will not be able to separate far enough from the $C_{60}$-dominated hybrid level. This, along with the fact that the overlap between $F$-$C_{60}$ and $F+C_{60}$ wavefunctions is large due to their non-vanishing amplitudes on both $F$ and $C_{60}$ regions, will favor their increased mixing in the resulting configuration of $F^-@C_{60}^{+(2p)}$.

It is true that the relatively weak hybridization in $F@C_{60}$ does not, as such, suggest too strong a reduction of screening to support a small enough up-shift of the F-dominant level so that the effect justifies the strong enhancement of hybridization as seen in $F^-@C_{60}^{+(2p)}$. But we argue that this is a limitation of the two-step model we use here to visualize the basic effect qualitatively, but not to quantify it. The real mechanism is a one-step adiabatic process where a $F^-$ anion is directly placed at the center of a $C_{60}^{+(2p)}$ cation so that the absence of an intermediate $F@C_{60}$ stage does not allow the states to relax. In effect, in the real system, the mixing and the transfer happen simultaneously between the same levels that hybridize. As pointed out, in the F atom, unlike other halogens, the valence $2p$ level has a higher binding energy than $2pC_{60}$, and that is the key here. This is so that with smaller screening in $F^-@C_{60}^{+(2p)}$ the partic-

ipant levels can still stay within a narrow energy proximity to strongly hybridize, while for $F^-@C_{60}^{+(H)}$ with full screening the levels cross but then separate out far enough to retain the weak hybridization. In contradistinction, the $np$ levels of Cl, Br and I being less bound than $2pC_{60}$ to begin with forbid this effect.

We note that previous structure calculations indeed predicted partial electron transfer from $C_{60}$ to F in $F@C_{60}$ with net charge on F to be -0.29 [35]. This study asserted no possibility of reverse transfer from F to $C_{60}$ and indicated ionization signals from hybridized $2pF$ orbitals. Separately, we have also computed the partial charge on F in the compound by using the density-derived electrostatic and chemical (DDEC) scheme [36] to find the charge of the F atom to be a fractional -0.43.

## B. Hybrid photoionization versus vacancy levels

Cross sections calculated in linear response time-dependent DFT for three systems $F@C_{60}$, $F^-@C_{60}^{+(H)}$, and $F^-@C_{60}^{+(2p)}$ are presented in Fig. 3 with panel 3(a) and 3(b), respectively, for emissions from the symmetric and the antisymmetric levels. It is seen that the lower-energy portions of all of the endofullerenes are fraught with resonances, primarily related to $C_{60}$ and are not there for free atoms (shown). The resonances result both from Auger and inter-Coulombic decay (ICD) processes [37]. Included in these cluster of resonances are the Auger-ICD hybridized decay processes as well [18]. In addition, note that the symmetric $F+C_{60}$ and antisymmetric $F$-$C_{60}$ level cross sections closely match, respectively, the antisymmetric $F^-$-$C_{60}^{+(H)}$ and symmetric $F^-$+$C_{60}^{+(H)}$ level cross sections. This is expected because their corresponding radial wavefunctions simply transpose the symmetry while exhibiting largely the same composition of mixing, as discussed in the previous subsection. The one noticeable difference is that the $F^-$-$C_{60}^{+(H)}$ cross section, due to this level's lower binding, opens at much lower photon energy than that of $F+C_{60}$.

### 1. Low energy spectra

Comparing all the hybrid level results with the $2p$ cross sections of free F and $F^-$, which practically overlap with each other, and with $2p$ of empty $C_{60}$ indicates plasmon driven enhancements and structures at lower photon energies [14, 17, 18, 38]. In the framework of interchannel coupling (IC) due to Fano [39], the correlation-modified matrix element of the photoionization of $X\pm C_{60}$ can be written perturbatively as [17, 24, 25],

$$\mathcal{M}_\pm(E) = \mathcal{D}_\pm(E)$$
$$+ \sum_{n\ell} \int dE' \frac{\langle \psi_{n\ell}(E') | \frac{1}{|\mathbf{r}_\pm - \mathbf{r}_{n\ell}|} | \psi_\pm(E) \rangle}{E - E'} \mathcal{D}_{n\ell}(E') \quad (3)$$



in which the single electron (uncorrelated) matrix element, that is the matrix element without $\delta V$ in Eq. (1), is

$$\mathcal{D}_\pm(E) = \langle ks(d)|z|\phi_\pm\rangle \qquad (4)$$

and $|\psi_{n\ell}\rangle$ in the IC integral is the (continuum) wavefunction of the $n\ell \to k\ell'$ channel. Taking the hybridization into account, the channel wavefunctions in Eq. (6) become

$$|\psi_\pm\rangle = \eta_\pm |\psi_{2pX}\rangle \pm \eta_\mp |\psi_{2pC_{60}}\rangle. \qquad (5)$$

Substituting Eqs. (5) into Eq. (6), and noting that the overlap between a pure X (X$^-$) bound state and a pure C$_{60}$ (C$_{60}^+$) bound state is negligible, since they occupy different regions of space, we separate the atomic and fullerene contributions to the integral to get the full (correlated) matrix element for X$\pm$C$_{60}$ and X$^-\pm$C$_{60}^+$ levels as,

$$\mathcal{M}_\pm(E) = \eta_\pm \mathcal{M}_{2pX(X^-)}(E) \pm \eta_\mp \mathcal{M}_{2pC_{60}(C_{60}^+)}(E) \qquad (6)$$

where the first and second terms, respectively, on the right hand side describes IC effects of atomic and fullerene ionization channels. Note that $\mathcal{M}_{2pX(X^-)}$ and $\mathcal{M}_{2pC_{60}(C_{60}^+)}$ are the IC matrix elements constructed *via* coupling respectively among *pure* atomic and fullerene channels.

Dominant fullerene characters of F$^-$+C$_{60}^{+(H)}$ and F-C$_{60}^{+(H)}$ wavefunctions (Fig. 2) ensure dominant fullerene IC effects in Eq. (6). This explains why the corresponding cross sections at lower energies almost follow the empty $2p$C$_{60}$ result in Fig. 3. For F$^-$+C$_{60}^{+(H)}$ and F+C$_{60}$ cross sections, on the other hand, while their dominant atomic character brings them close to the free $2p$F (F$^-$) results at higher energies, their enhancements at lower energies are notable. This enhancement is due to an IC driven mechanism discussed previously [38] that siphons off giant-size strengths from the fullerene plasmonic emission even though the fullerene admixture of these hybrid wavefunctions is quite small. However, note a rather dramatic difference between the symmetric F$^-$+C$_{60}^{+(2p)}$ and F$^-$+C$_{60}^{+(H)}$ results in Fig. 3(a) over, in fact, the entire energy range shown. This owes to the increased hybridization process in the F$^-$+C$_{60}^{+(2p)}$ discussed above as the main finding of this study. The process subsequently yields comparable atomic and fullerene IC effects in Eq. (6) including a comparable contribution of their coherence. On the other hand, there is not-so-dramatic differences between antisymmetric F$^-$-C$_{60}^{+(2p)}$ and F$^-$-C$_{60}^{+(H)}$ results in Fig. 3(b). In sum, we find a most interesting and unexpected phenomenology; the symmetric case, Fig. 3(a), is very strongly dependent upon the location of the C$_{60}$ vacancy, but the antisymmetric case is not. This must mean that there are cancellations in the matrix elements in the antisymmetric case that are not present for the symmetric case.

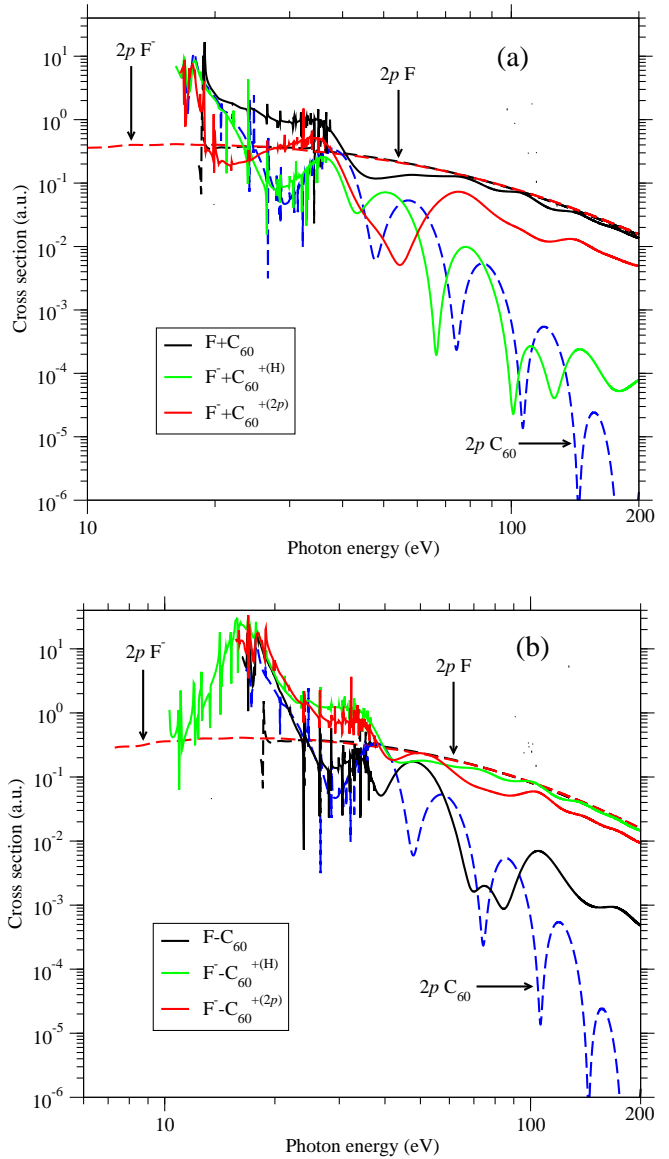

FIG. 3. (Color online) Photoionization cross sections of the symmetric (a) and antisymmetric (b) hybrid levels of F@C$_{60}$, F$^-$@C$_{60}^{+(H)}$, and F$^-$@C$_{60}^{+(2p)}$. Cross sections of $2p$F (F$^-$) and $2p$C$_{60}$ are also presented.

#### 2. Higher energy spectra

As the plasmonic effect weakens with increasing energy, oscillatory modulations at varied degrees of prominence show up in the cross sections. These oscillations are a consequence of a well-known interference mechanism [40] due to the cavity structure of C$_{60}$ which was modeled earlier in detail in Ref. [41]. At such high energies the IC in Eq. (6) is no longer important, but the hybridization remains, so that Eqs. (6) simplify to

$$\mathcal{D}_\pm(E) = \eta_\pm \mathcal{D}_{2pX(X^-)}(E) \pm \eta_\mp \mathcal{D}_{2pC_{60}(C_{60}^+)}(E). \qquad (7)$$

The interference model is based on the following mechanism. The matrix element $\mathcal{D}_{\pm}$ can generally be separated into two components, one arising from the atomic region and other from the $C_{60}$ shell region, and can be written down, respectively, as follows [41]

$$\mathcal{D}_X \sim \mathcal{D}^{\text{atom}}(k)$$
$$+ A^{\text{refl}}(k)\left[e^{-ikD_o}e^{-iV_0\frac{2\Delta}{k}} - e^{-ikD_i}\right] \quad (8a)$$

$$\mathcal{D}_C \sim A^{\text{shell}}(k)e^{-i\frac{V_0}{k}}\left[a_i e^{-ikR_i} - a_o e^{-ikR_o}\right], \quad (8b)$$

where the photoelectron momentum $k = \sqrt{2(E - \epsilon_{\pm})}$ in atomic units, $a_i$ and $a_o$ are the values of $\phi_{\pm}$ at the inner and outer radii $R_i$ and $R_o$ of $C_{60}$, and $V_0$ is the average depth of the shell potential. In Eq. (8a) $\mathcal{D}^{\text{atom}}$ represents the direct ionization amplitude from the atomic region. The second term in this equation embodies the reflection of this outgoing photoelectron wave from both inner and outer surfaces of the shell. Quantitatively, this induces oscillations as a function of the photoelectron momentum with amplitude $A^{\text{refl}}$ and frequencies related to $D_i$ and $D_o$, the inner and outer diameters of the shell. The direct and the reflected parts coherently interfere in the cross section.

Since $A^{\text{refl}}$ is proportional to $\mathcal{D}^{\text{atom}}$, the larger the atomic component of a hybrid wavefunction, the stronger is the reflection and the higher is the chances that the oscillations occur about the free atom (ion) result. This is exactly what is seen for the high energy cross section of the F-dominant levels of F@$C_{60}$ and F$^-$@$C_{60}^{+(H)}$ that follow the $2p$F result in Fig. 3. On the other hand, Eq. (8b), resulting from the overlap integral in the shell region, produces two localized emissions in the vicinities of the shell edges, where the available ionizing forces maximize due to rapid variations of the shell potential there. Such a diffraction-type effect translates into another oscillation in frequencies related to $R_i$ and $R_o$. This part will also add to the coherent interference in the cross section and will dominate if a hybrid level has a stronger $C_{60}$ character, like for the $C_{60}$-dominant levels of F@$C_{60}$ and F$^-$@$C_{60}^{+(H)}$ in Fig. 3. Indeed, the oscillation structures in the cross section of these hybrid levels intensify at higher energies while the average value fall lower than free $2p$F as exhibited in Fig. 3.

As already noted above, for the F$^-$@$C_{60}^{+(2p)}$ hybrids, due to their almost comparable share of atom-$C_{60}$ character, the differences in the corresponding cross section shapes arise from the interference among strong direct atomic, reflective and relatively comparable diffractive emissions. As pointed out above, the particularly strong differences between the high-energy results of symmetric F$^-$+$C_{60}^{+(2p)}$ and F$^-$+$C_{60}^{+(H)}$ and relatively weaker differences between antisymmetric F$^-$-$C_{60}^{+(2p)}$ and F$^-$-$C_{60}^{+(H)}$ are noted in Fig. 3. These differences further identify the importance of hybrid versus pure vacancies in F$^-$@$C_{60}^+$ in the spectroscopic details of higher energy emission structure and properties.

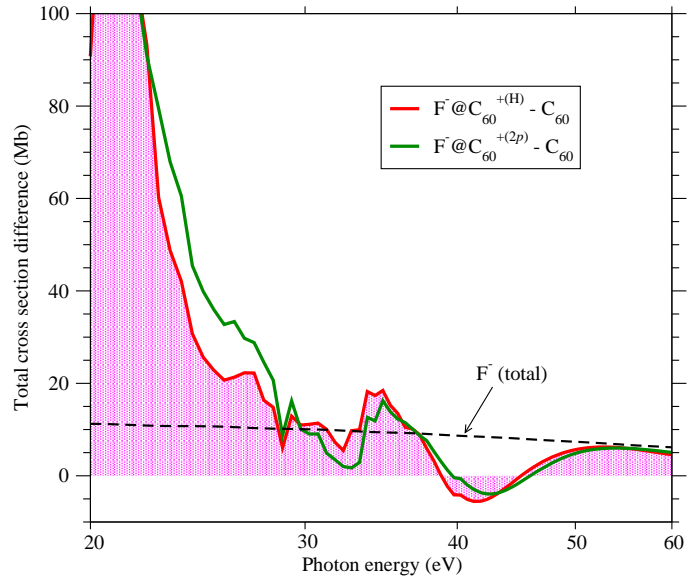

FIG. 4. (Color online) The magnitude differences of the total photoionization cross sections of F$^-$@$C_{60}^{+(H)}$ and F$^-$@$C_{60}^{+(2p)}$ from that of the empty $C_{60}$. Some smoothing is included. The total cross sections of F$^-$ is presented for comparison.

## C. Experimental possibilities

Fig. 4 presents the results of the subtraction of total $C_{60}$ cross section from the total cross sections of the compound in two configurations F$^-$-$C_{60}^{+(2p)}$ and F$^-$-$C_{60}^{+(H)}$. The deviation of the results from the total cross section of free F$^-$ ion (shown) captures the net effect of the atom-fullerene coherence following the electron transfer. As evident, constructive interferences from this coherence is seen near the $C_{60}$ giant plasmon resonance energies. Surprisingly, however, this effect at the higher energy (weaker) resonance around 40 eV is found rather destructive. Thus, such coherence effects can be directly probed by total photoionization cross section measurements by, for instance, the techniques utilized earlier based on detection and analysis of photoions [8–10]. In fact, such measurements can be extended for other halogen endofullerenes as well. Since the results in Fig. 4 also accounts for the differences between vacancy positions, the comparison with measurements may further shed light on the possible configuration of the molecule in its ground state.

Another method to explore the situation experimentally is to use one laser pulse to excite F@$C_{60}$ by inducing an electron transfer and another pulse to ionize the hybrid levels to probe excited state cross sections. Such two-color two-step photoionization measurement techniques have been used earlier [42–44]. Based on this technique, an electron can be selectively promoted from a chosen $C_{60}$ level, namely $C_{60}$ HOMO or a hybrid level, to form metastable configurations. The comparison between spectra for photoelectrons driven by the subsequent second laser allows access to ionization properties as a function of the vacancy level.

Synthesis of halogen endofullerenes in vapor-phase has not been reported yet. But this may be possible by, for instance, halogen ion implantation techniques as adopted for other systems [45]. Perhaps because of the noncovalent interactions of these molecules with the environment their photoionization response will remain largely intact in some stable salt form, or in solutions or as thin films.

## IV.  CONCLUSIONS

We previously studied [24, 25] the photoionization of various halogen-atomic (Cl, Br and I) endofullerene molecules with a $C_{60}$ electron relocated to the halogen atom likely forming metastable configurations. Those studies predicted that the effect of atom-$C_{60}$ hybridization on both ground state and photoemission properties is practically insensitive to the location of the vacancy level in $C_{60}$. In the current investigation we find a strong exception to this rule for such a metastable system $F^-@C_{60}^+$ made of the remaining halogen atom F. This exception reveals a novel effect based on the relative positions of the levels that hybridize in the compound. It is found that since, unlike to other halogens, the participating F level is more bound than the partner $C_{60}$ level, the increased screening due to the transfer of the electron results in a level crossing mechanism. Due to this mechanism, the degree of hybridization is significantly altered based on the vacancy creation in the hybrid-active versus a pure $C_{60}$ level. Consequently the hybrid photoemission properties become a very sensitive function of the vacancy position. The calculations are performed in a linear-response time-dependent density functional scheme, as was also employed in previous studies. Even though the effect is unique for $F^-@C_{60}^+$ for the halogen series, the precondition of such level-crossing effect can easily be satisfied for some of the other atoms, or small molecules, or metal cluster trapped in $C_{60}$ or even, in general, for different choices of fullerenes itself, given that the atom-fullerene hybridization is a ubiquitous phenomenon in a host of these materials. In other words, we expect that this is a very general phenomenon.

In our calculations, the $C_{60}$ ion core is smeared in a jellium spherical shell that freezes lattice vibrations. Effectively, this mimics results at absolute zero sample temperature. However, it was shown earlier [27] that finite oven temperature effects, such as, the electron-phonon coupling [46] and fluctuation of the shape around the shape at absolute zero [47], approximately needed an extra width less than 1 eV to compare with measurements for $C_{60}$. This width is smaller than an energy resolution of about 5 eV required to measure broad structures in Figs. 3 and 4. Thus, while the thermal vibration will likely wash out the autoionizing spikes, it would not qualitatively alter the key results of this study. Furthermore, possible temperature driven oscillations [48] of the confined atom may broaden and lower some narrow structures in the signal, but most of the features in the current results being broad enough will survive. As an example, the measurement of the confinement induced structures in Xe@$C_{60}$ was possible in the laboratory at a temperature well above absolute zero [9, 10].

## ACKNOWLEDGMENTS

The research is supported by the US National Science Foundation Grant No. PHY-1806206 (HSC) and the US Department of Energy, Office of Science, Basic Energy Sciences, under award DE-FG02-03ER15428 (STM).